\documentclass[jgrga]{AGUTeX}

\usepackage{color}
\usepackage{lineno}

 \usepackage[xdvi]{graphicx}

\setkeys{Gin}{draft=false}
\def\rv{{\mathbf r}}   
\def\kv{{\mathbf k}}  
\def\xv{{\mathbf x}}  

\authorrunninghead{BADULIN \& ZAKHAROV}

\titlerunninghead{Ocean swell within the kinetic equation}

\authoraddr{S.~I.~Badulin,
P.~P.~Shirshov Institute of Oceanology of the Russian Academy of Science, 36, Nakhimovsky pr., 117997, Moscow, Russia,  (badulin@ioran.ru)}

\authoraddr{V.~E.~Zakharov,
University of Arizona, Tuscon, USA\\
 Novosibirsk State University, Russia\\
P.N. Lebedev Physical Institute of Russian Academy of Sciences\\
Waves and Solitons LLC, Phoenix, Arizona, USA}
\graphicspath{{Fig/}}

\begin{document}

\title{Ocean swell within the kinetic equation for water waves}%
\authors{S. I. Badulin\altaffilmark{1-2}, V. E. Zakharov
\altaffilmark{1-5}}
\altaffiltext{1}{P.~P.~Shirshov Institute of Oceanology of the Russian Academy of Science, 36, Nakhimovsky pr., 117997, Moscow, Russia}
\altaffiltext{2}{Laboratory of Nonlinear Wave Processes, Novosibirsk State University, Russia}
\altaffiltext{3}{University of Arizona, Tuscon, USA}
\altaffiltext{4}{P.N. Lebedev Physical Institute of Russian Academy of Sciences}
\altaffiltext{5}{Waves and Solitons LLC, Phoenix, Arizona, USA}

\begin{abstract}
Effects of wave-wave interactions on  ocean swell are studied. Results of extensive simulations of swell evolution within the duration-limited setup for the kinetic Hasselmann equation at long times up to $10^6$ seconds  are presented. Basic solutions of the theory of weak turbulence, the so-called Kolmogorov-Zakharov solutions, are shown to be relevant to the results of the simulations. Features of self-similarity of wave spectra are detailed and their impact on methods of ocean swell monitoring are discussed. Essential drop of wave energy (wave height) due to wave-wave interactions is found to be pronounced at initial stages of swell evolution (of order of 1000 km for typical parameters of the ocean swell). At longer times wave-wave interactions are responsible for a universal angular distribution of wave spectra in a wide range of initial conditions.
\end{abstract}

\begin{article}
\section{Physical models of ocean swell}
Ocean swell is an important constituent of  the field of surface gravity waves in the sea and, more generally, of the sea environment as a whole. Swell is usually defined as a fraction of wave field that does not depend (or depends slightly) on local wind. Being generated in relatively small stormy areas these waves can propagate at long distances of many thousand miles, thus, influencing vast ocean stretches. For example, swell from Roaring Forties in the Southern Ocean  can traverse  the Pacifica  and reach distant shores of California and Kamchatka. Predicting of swell as a part of surface wave forecast remains  a burning problem for maritime safety and marine engineering.

Pioneering works by \citet{MunkSwell1963,Snodgrass1966} discovered a rich physics of the phenomenon  and gave first examples of accurate measurements of magnitudes, periods and directional spreading of swell. Both articles \citep{MunkSwell1963,Snodgrass1966} contain thorough discussions of physical background of  swell generation, attenuation and interaction with other types of ocean motions. Nonlinear wave-wave interactions have been  sketched in these articles as a novelty introduced by the milestone papers by \citet{Phillips60} and \citet{Hass1962}. A possible important role of these interactions at high swell heights for relatively short time (fetch) of evolution has been outlined and estimated. The first estimates of the observed rates of swell attenuation have been carried out by \citet{Snodgrass1966} based on observation at near-shore stations. Their characteristic scale (e-folding) about $4000$ km is consistent with some today results of the satellite tracking of swell \citep{Ardhuinetal2009,Ardhuin2010dissipation,Jiang2016} and with treatment of these results within the model of swell attenuation due to coupling with turbulent atmospheric layer. Alternative  model of turbulent wave flow attenuation \citep{Babanin2006turb} predicts quite different algebraic law and stronger swell attenuation at shorter distances from the swell source \citep{YoungBabZig2013jpo}. It should be stressed that all the mentioned models treat swell as a quasi-monochromatic wave and, thus, ignore nonlinear interactions of the swell harmonics themselves and the swell coupling with locally generated wind-wave background. The latter effect can be essential as observations and simulations clearly show \citep[][and refs. therein]{Young2006,BKRZ2008Brest}.

Today, linear models of swell are dominating. The swell is generally considered as a superposition of harmonics that do not interact to each other and, thus, can be described by the well-known methods of the linear theory of  waves. Effects of weak dissipation, refraction by underwater relief and islands, sphericity of the Earth can be accounted for as well. Some features of the observed swell can be related  to such models. For example, the observed linear growth of the swell frequency in a site can be explained as an  effect of dispersion of a linear wave packet at long time. The linear models of swell dissipation lead, evidently, to exponential laws of the swell energy (height) attenuation and to attempts to estimate the e-folding scale from available experimental data \citep[e.g.][]{Snodgrass1966,Jiang2016}.

Synthetic aperture radars (SAR) allow for  spatial resolution up to tens meters and, thus,  for detecting relatively long swell waves of a few hundred meters wavelength along their thousand miles tracks \citep[e.g.][]{Ardhuin2010dissipation,YoungBabZig2013jpo}. Satellite altimeters  measure wave height averaged over a snapshot of a few square kilometers that is adequate to the today methods of statistical description of waves in the research and application models and also can be used for the swell tracking in combination with other tools \citep[e.g. wave models as in ][]{Jiang2016}. Re-tracking of swell allows, first, for relating the swell events with their probable sources -- stormy areas. Secondly, the swell transformation  gives a clue to estimating effects of other motions of the ocean \citep[e.g.][]{ChenEtal2002}.  Such work requires adequate physical models of swell propagation and transformation as far as a number of parameters of sea environment remains beyond  control.

The linear treatment of swell is not able to explain observed features of swell. First, the observed swell spectra exhibit frequency downshifting which is not predicted by linear or weakly nonlinear models of wave guide evolution \citep[e.g. data of][and comments on these data by \citet{HendersonSegur2013}]{Snodgrass1966}. Secondly, these spectra show clearly invariance of their shaping that is unlikely to be appear in linear dispersive wave system. All the noted features are common for wave spectra described by the kinetic equation for water waves, the so-called \citet{Hass1962} equation.

In this paper we present results of extensive  simulations of ocean swell within the Hasselmann equation. The simplest duration-limited setup has been chosen to obtain numerical solutions for the duration up to $10^6$ seconds ($11.5$ days) for typical parameters of ocean swell (wavelengths $150-400$ meters, initial significant heights $3-10$ meters).

We analyze the simulation results from the viewpoint of the theory of weak turbulence \citep{ZakhFalLvov92}. The slowly evolving swell solutions appear to be quite close to the stationary milestone Kolmogorov-Zakharov solutions for water waves \citep{ZakhFil66,ZakhZasl82a} in a frequency range and give estimates for the fundamental constants of the theory. We give a short theoretical introduction and present these estimates in the next section. In sect.~3 we relate results of simulations with properties of the self-similar solutions of the kinetic equation. \citet{Zaslavsky2000} was the first who presented the self-similar solutions for swell assuming the angular narrowness of the swell spectra and gave a set of explicit analytical expressions for the swell. In fact, more general consideration in the spirit of \citet{BPRZ2002,BPRZ2005} leads to important findings and to a number of questions with no reference to the assumption of angular narrowness.

We demonstrate the well-known fact that is usually ignored: the power-law  swell attenuation within the conservative kinetic equation. We show that it does not contradict to  results of observations mentioned above. We also fix a remarkable feature of collapsing the swell spectra to an angular distribution that depends weakly on initial angular spreading. Such universality can be of great value for modelling swell and developing methods of its monitoring.

The paper is finalized by discussion of the effect wave-wave interactions on swell attenuation. We consider a case where long-time swell evolution is perturbed by light ($5$ m/s) wind which direction is varying with inertial period. We show that the mechanism of inverse cascading due to weakly nonlinear wave-wave interactions provides an  absorption of energy of locally generated short wind waves and an effective `feeding' of long sea swell \citep{BKRZ2008Brest}. This effect remains beyond the attention of and cannot be treated within the models of swell attenuation mentioned above. Meanwhile, recent observations of swell from space show definitely possibility of swell amplification \citep[`negative' dissipation in words of][]{Jiang2016}.

\section{Solutions for ocean swell}
\subsection{The Kolmogorov-Zakharov solutions}
In this section we reproduce previously reported theoretical results on evolution of swell as a random field of  weakly interacting wave harmonics. We follow the statistical theory of wind-driven seas \citep{Zakharov99} extending this approach to the sea swell which description within this approach is usually considered as questionable. A random wave field is described by the kinetic equation derived by Klauss Hasselmann in early sixties \citep{Hass1962} for deep water waves in absence of dissipation and external forcing
\begin{linenomath*}
\begin{equation}
\label{eq:Kinfull} \frac{\partial N_{\kv}}{\partial t} +
\nabla_{\kv} \omega_{\kv} {\nabla_\rv N_{\kv}} = S_{nl}.
\end{equation}
\end{linenomath*}
Equation (\ref{eq:Kinfull}) is written for the spectral density of wave action $N(\kv,\xv,t)=E(\kv,\xv,t)/\omega(\kv)$ ($E(\kv,\xv,t)$ is wave energy spectrum and wave frequency obeys linear dispersion relation $\omega=\sqrt{g|\kv|}$). Subscripts for $\nabla$ corresponds to the two-dimensional gradient operator in the corresponding space of coordinates $\xv$ and wavevectors $\kv$ (i.e. $\nabla_{\xv}=(\partial/\partial x,\partial/\partial y)$).

The right-hand term $S_{nl}$ describes the effect of wave-wave resonant interactions and can be written in explicit form \citep[see][for collection of formulas]{BPRZ2005}. The cumbersome term $S_{nl}$ causes a number of problems for wave modelling where (\ref{eq:Kinfull}) is extensively used. Nevertheless, for deep water case one has a key property of the term homogeneity
\begin{linenomath*}
\begin{equation}\label{eq:col_hom_k}
    S_{nl}[\kappa \kv, \nu N_\kv] = \kappa^{19/2} \nu^3 S_{nl}[ \kv,  N_\kv].
\end{equation}
\end{linenomath*}
that helps a lot in getting important analytical results.
Stretching in $\kappa$ times in wave scale or in $\nu $ times in wave action ($\kappa,\,\nu$ are positive) leads to simple re-scaling of the collision term $S_{nl}$. This important property gives a clue for constructing power-law stationary solutions of the kinetic equation, i.e. solutions for the equation
\begin{linenomath*}
\begin{equation}\label{eq:Snl0}
  S_{nl}=0.
\end{equation}
\end{linenomath*}
Two isotropic stationary solutions of (\ref{eq:Snl0}) correspond to constant fluxes of wave energy and action in wave scales. The direct cascade solution \citep{ZakhFil66} in terms of frequency spectrum of energy
\begin{linenomath*}
 \begin{equation}\label{eq:Kolmogorov_direct}
   E^{(1)}(\omega,\theta) = \frac{C_p}{2} \frac{P^{1/3}g^{4/3}}{\omega^{4}}
\end{equation}
\end{linenomath*}
introduces the basic Kolmogorov constant $C_p$ and describes the energy transfer to infinitely short waves with constant flux $P$. The wave action transfer to opposite direction of long waves is described by the inverse cascade solution \citep{ZakhZasl82a}
\begin{linenomath*}
\begin{equation}\label{eq:Kolmogorov_inverse}
  E^{(2)}(\omega,\theta) = \frac{C_q}{2} \frac{Q^{1/3}g^{4/3}}{\omega^{11/3}}
\end{equation}
\end{linenomath*}
with wave action flux $Q$ and another Kolmogorov's constant $C_q$.

An approximate weakly anisotropic Kolmogorov-Zakharov solution has been obtained by \citet{Katz_Kontor74} as an extension of (\ref{eq:Kolmogorov_direct})
\begin{linenomath*}
 \begin{equation}\label{eq:Kolmogorov_aniso}
   E^{(3)}(\omega,\theta) = \frac{P^{1/3}g^{4/3}}{2\omega^{4}}\left(C_p + C_m\frac{gM}{\omega P}\cos\theta +\ldots\right).
\end{equation}
\end{linenomath*}
(\ref{eq:Kolmogorov_aniso}) associates the wave spectrum anisotropy with the constant spectral flux of wave momentum $M$ and the so-called second Kolmogorov constant  $C_m$. As it is seen from (\ref{eq:Kolmogorov_aniso}) the solution  anisotropy vanishes as $\omega\to \infty$: wave spectra become isotropic for short waves. A harmony of the above set of the KZ solutions can be treated naturally within the dimensional approach (\ref{eq:Kolmogorov_direct}--\ref{eq:Kolmogorov_aniso}): they are just particular cases of solutions of the form
\begin{linenomath*}
\begin{equation}\label{eq:KZgeneral}
  E^{(KZ)}(\omega) = \frac{P^{1/3}g^{4/3}}{\omega^{4}}G(\omega Q/P,gM/(\omega P),\theta)
\end{equation}
\end{linenomath*}
where $G$ is a function of dimensional arguments scaled by spectral fluxes of energy $P$, wave action $Q$ and wave momentum $M$.

Originally, solutions (\ref{eq:Kolmogorov_direct}--\ref{eq:Kolmogorov_aniso}) have been derived in quite sophisticated and cumbersome ways. Later on, simpler and more physically transparent  approaches have been presented for generalization the set of these solutions and looking for their higher-order anisotropic extensions \citep{ZakhPush1999,Balk2000,PRZ2003,PRZ2004,BPRZ2005,Zakharov2010Scr}. In particular, the extension of the series in the right-hand side of (\ref{eq:Kolmogorov_aniso}) predicts the next term proportional to $\cos2\theta/\omega^2$ that is the second angular harmonics.

Swell solutions evolve slowly with time and, thus, give a good opportunity for discussing features of the KZ solutions (or, alternatively, the KZ solutions can be used as a benchmark for the swell studies). One of the key points of this discussion is the question of uniqueness of the swell solutions. It can be treated in the context of general KZ solutions (\ref{eq:KZgeneral}). While the principal terms of the general Kolomogorov-Zakharov solution corresponding to (\ref{eq:Kolmogorov_direct}--\ref{eq:Kolmogorov_aniso}) have clear physical meaning of total fluxes of wave action (\ref{eq:Kolmogorov_inverse}), energy (\ref{eq:Kolmogorov_direct}) and momentum (\ref{eq:Kolmogorov_aniso}) this is not the case of the higher-order terms. The link of these additional terms with inherent properties of the collision integral $S_{nl}$ or/and with specific initial conditions is not a trivial point. In this paper we give the very preliminary answer to this intriguing question.

\subsection{Self-similar solutions of the kinetic equation}
The homogeneity property (\ref{eq:col_hom_k}) is extremely useful for studies of non-stationary (inhomogeneous) solutions of the kinetic equation. Approximate self-similar solutions for reference cases of duration- and fetch-limited development of wave field can be obtained under assumption of  dominance of the wave-wave interaction term $S_{nl}$ \citep{PRZ2003,Zakharov2005NPG,BPRZ2005,ZakharovBadulin2011DAN}. These solutions have forms of the so-called incomplete or the second type self-similarity \citep[e.g.][]{Barenblattbook79}. In terms of frequency-angle dependencies of wave action spectra one has for the duration- and fetch-limited cases correspondingly \citep{BPRZ2005,BBRZ2007,Badulin2015univ}
\begin{linenomath*}
\begin{eqnarray}\label{eq:DurSS}
  N(\omega,\theta, \tau) & = & a_\tau \tau^{p_\tau} \Phi_{p_\tau}(\xi,\theta)\\
  \label{eq:FetchSS}
N(\omega,\theta, \chi) &= & a_\chi \chi^{p_{\chi}} \Phi_{p_\chi}(\zeta,\theta)
\end{eqnarray}
\end{linenomath*}
with dimensionless time $\tau $ and fetch $\chi$
\begin{linenomath*}
\begin{equation}\label{eq:timefetchDef}
  \tau=t/t_0; \qquad
  \chi=x/x_0.
\end{equation}
\end{linenomath*}
Homogeneity properties (\ref{eq:col_hom_k}) dictates `magic relations' \citep[in words of][]{PushZakh2015arXiv,Pushkarev2016} between dimensionless exponents $p_\tau,\, q_\tau$ and $p_\chi,\, q_\chi$
\begin{linenomath*}
\begin{equation}\label{eq:magic}
  p_\tau=\frac{9q_\tau-1}{2}; \qquad p_\chi=\frac{10q_\chi-1}{2}.
\end{equation}
\end{linenomath*}
Dimensionless arguments of shaping functions $\Phi_{p_\tau}(\xi),\,\Phi_{p_\chi}(\zeta)$ contain free scaling parameters $b_\tau,\,b_\chi$
\begin{linenomath*}
\begin{equation}\label{eq:SSargs}
  \xi=b_\tau \omega^2 \tau^{-2q_\tau}; \qquad \zeta=b_\chi \omega^2 \chi^{-2q_\chi}.
\end{equation}
\end{linenomath*}
Additional `magic relations' coming from homogeneity property (\ref{eq:col_hom_k}) fix a link between amplitude scales $a_\tau,\,a_\chi$ and the bandwidth scales  $b_\tau,\,b_\chi$ of the self-similar solutions (\ref{eq:DurSS},\ref{eq:FetchSS})
\begin{linenomath*}
\begin{equation}\label{eq:a2b}
  a_\tau=b_\tau^{19/4};\qquad a_\chi=b_\chi^{5/2}.
\end{equation}
\end{linenomath*}
Thus, `magic relations' (\ref{eq:magic},\ref{eq:a2b}) reduce number of free parameters of the self-similar solutions (\ref{eq:DurSS},\ref{eq:FetchSS}) from four (two exponents and two coefficients) to two  only: a dimensionless exponent $p_\tau\,(p_\chi)$ and an amplitude of the solution $a_\tau\,(a_\chi)$.

The shaping functions $\Phi(\xi),\,\Phi(\zeta)$ in (\ref{eq:DurSS},\ref{eq:FetchSS}) require solution of a boundary problem for an integro-differential equation in self-similar variable $\xi $ (or $\zeta$ for fetch-limited case) and angle $\theta$ \citep[see sect.~5.2][for details]{BPRZ2005}. Simulations \citep[e.g.][]{BBZR2008} show remarkable features of the shaping functions $\Phi(\xi),\,\Phi(\zeta)$. First, numerical solutions generally show relatively narrow angular distributions for $\Phi_{p_\tau}(\xi),\,\Phi_{p_\chi}(\zeta)$ with a single pronounced maximum near a spectral peak frequency $\omega_p$.  It implies that the only one (or very few) of an infinite series of eigenfunctions  of the boundary problem for the shaping functions $\Phi_{p_\tau}(\xi),\,\Phi_{p_\chi}(\zeta)$ contributes into wave spectra evolution in a wide range of initial and external forcing conditions. This treatment of the heavily nonlinear boundary problem in terms of a composition of eigenfunctions is possible in this case as demonstrated by \citet{ZakhPush1999}. Two-lobe patterns are observed at higher frequencies ($\omega > 2\omega_p$) in some cases as local maxima at oblique directions or as a `shoulder' in wave frequency spectra. Their appearance is generally discussed as an effect of wind on wave generation \citep[e.g.][]{BottemaVledder2008,BottemaVledder2009}.

Secondly, an important property of \emph{spectral shape invariance} \citep[terminology of][]{Hass_ross_muller_sell76} or \emph{the spectra quasi-universality} \citep[in words of][]{BPRZ2005} is widely discussed both for experimentally observed and simulated wave spectra. This  invariance does not suppose a point-by-point coincidence of properly normalized spectral shapes. Proximity of integrals of the shape functions $\Phi_{p_\tau},\, \Phi_{p_\chi}$ in a range of wave growth rates $p_\tau,\, p_\chi$ appears to be sufficient for formulating a remarkable universal relationship for parameters of self-similar solutions (\ref{eq:DurSS},\ref{eq:FetchSS})
\begin{linenomath*}
\begin{equation}\label{eq:univ}
  \mu^4\nu=\alpha_0^3.
\end{equation}
\end{linenomath*}
Here wave steepness $\mu$ is estimated from total wave energy $E$ and spectral peak frequency $\omega_p$
\begin{linenomath*}
\begin{equation}\label{def:steepness}
  \mu=\frac{E^{1/2}\omega_p^2}{g}.
\end{equation}
\end{linenomath*}
The `number of waves' $\nu$ in a spatially homogeneous wind sea (i.e. for duration-limited wave growth) is defined as follows:
\begin{linenomath*}
\begin{equation}\label{eq:nu}
  \nu=\omega_p t.
\end{equation}
\end{linenomath*}
For spatial (fetch-limited) wave growth  the coefficient of proportionality $C_f$  in the equivalent expression $\nu=C_f |\kv_p|x$ ($\kv_p$ being the wavevector of the spectral peak) is close to the ratio between the  phase and  group velocities $C_{ph}/C_{g}=2$. A universal constant $\alpha_0 \approx 0.7$ is a counterpart of the constants $C_p,\,C_q $ of the stationary Kolmogorov-Zakharov solutions (\ref{eq:Kolmogorov_direct},\ref{eq:Kolmogorov_inverse}) and has a similar physical meaning of a ratio between wave energy and the energy spectral flux (in power $1/3$). A remarkable feature of the universal wave growth law (\ref{eq:univ}) is its independence on wind speed. This wind-free paradigm based on intrinsic scaling of wave development is capable to become a useful tool of analysis of both wind-wave growth and wind-free ocean swell \citep{Badulin2015univ}.

\subsection{Self-similarity of swell solutions}
The self-similar solution for swell is just a  member of a family of solutions (\ref{eq:DurSS},\ref{eq:FetchSS}) with particular values of temporal or spatial rates
\begin{linenomath*}
\begin{eqnarray}\label{eq:swelltau}
  p_\tau=1/11; & \qquad q_\tau=1/11\\
  \label{eq:swellchi}
  p_\chi=1/12; & \qquad q_\chi=1/12
\end{eqnarray}
\end{linenomath*}
Exponents (\ref{eq:swelltau},\ref{eq:swellchi}) provide conservation of the total wave action for its evolution in time (duration-limited setup) or in space (fetch-limited)
\begin{linenomath*}
\begin{equation}\label{eq:ActionConst}
N=\int_0^\infty N(\omega,\theta) d\omega d\theta = \rm{const}
\end{equation}
\end{linenomath*}
On the contrary, total energy and wave momentum are formal constants of motion  of the Hasselmann equation  and decay with time $t$ or fetch $x$
\begin{linenomath*}
\begin{eqnarray}\label{eq:EnMom1}
  E \sim t^{-1/11}; & \qquad M \sim t^{-2/11}\\
    E \sim x^{-1/12}; & \qquad M \sim x^{-2/12}
    \label{eq:EnMom2}
\end{eqnarray}
\end{linenomath*}
The swell decay  (\ref{eq:EnMom1},\ref{eq:EnMom2}) reflects a basic feature of the kinetic equation for water waves: energy and momentum are not conserved \citep[see][and refs. herein]{ZakhFalLvov92,PRZ2003}. The wave action is the only true integral of the kinetic equation (\ref{eq:Kinfull}).

The swell solution  manifests another general feature of evolving wave spectra: the downshifting of the spectral peak frequency (or other characteristic frequency), i.e.
\begin{linenomath*}
\begin{equation}\label{eq:downshift}
  \omega_p \sim t^{-1/11}; \qquad \omega_p \sim x^{-1/12}.
\end{equation}
\end{linenomath*}
The universal law of wave evolution (\ref{eq:univ}) is, evidently, valid for the self-similar swell solution as well with a minor difference in the value of the constant $\alpha_0$. As soon as this constant  is expressed in terms of the integrals of the shape functions $\Phi_\tau,\,\Phi_\chi$ and the swell spectrum shape differs essentially from ones of the growing wind seas this constant appears to be less than $\alpha_0$ of the growing wind seas.

The theoretical facts presented above gives a background for analysis of results of our simulations.

\section{Swell simulations}
\subsection{Simulation setup}
Simulations  of ocean swell  require special care. First of all, calculations for quite long time (up to $10^6$ seconds in our case) should be accurate enough in order to fix relatively slow evolution of solutions and, thus, be able to relate results with the theoretical background presented above. We used the approach of our previous papers \citep{BPRZ2002,BPRZ2005,BBRZ2007,BBZR2008}.

Duration-limited evolution has been simulated with the code based on WRT algorithm \citep{Webb78,Tracy82} for different initial conditions (spatial spectra at $t=0$). Frequency resolution for log-spaced grid has been set to $(\omega_{n+1}-\omega_{n})/\omega_n=1.03128266$. It corresponds to $128$ grid point in frequency range $0.02-1$ Hz (approximately $1.5$ to $3850$ meters wave length). Thus, we used very fine  frequency resolution in a wide range of wave scales \citep[cf.][]{Benoit_Gagnaire_2007,GBB2011} to trace rather slow evolution of swell. Standard angular resolution $\Delta \theta = 10^\circ$ has been taken as adequate to goals of our study.

Initial conditions were similar in all series of simulations: wave action spectral density in a box of frequencies and angles was slightly ($5\% $ of magnitude) modulated  in  angles and had low (six orders less) background outside this box:
\begin{linenomath*}
\begin{equation}\label{eq:ini_spectra}
  N(\kv)=\left\{
\begin{array}{l}
  N_0(1+0.05\cos(\theta^2/2)),\,
   |\theta| < \Theta/2, \,\omega_l<\omega <\omega_h \\
  10^{-6}N_0,\qquad \rm{otherwise}
  \end{array}\right.
\end{equation}
\end{linenomath*}
 The modulations have been set in order to stimulate wave-wave interactions as well as the collision integral $S_{nl}$ vanishes for $N(\kv)=\rm{const}$. In contrast to wind waves where wind speed is an essential physical parameter that provides a useful physical scale, the swell evolution is determined by initial conditions only, i.e. by $N_0$ and $\Theta$ in the setup (\ref{eq:ini_spectra}).

Dissipation was absent in the runs. Free boundary conditions were applied at the high-frequency end of the domain of calculations: generally, short-term oscillations of the spectrum tail do not lead to instability. Calculations with a hyper-dissipation \citep[e.g.][]{PRZ2003} or a diagnostic tail at the high-frequency range of the spectrum \citep[][]{GagnaireEtal2010} do not affect  results  even quantitatively as compared with our options of simulations.

Totally, more than $30$ runs have been carried out for different initial conditions for duration $10^6$ seconds. Below we focus on the series of Table~\ref{table1}  where initial wave heights were fixed (within $2\%$) and angular spreading varied from very narrow $\Theta =30^\circ$ to almost isotropic $\Theta=330^\circ$ (\ref{eq:ini_spectra}). The frequency range of the initial perturbations was  $0.1-0.4$Hz.
\begin{table}
\caption{Initial parameters of simulation series}
\centering
\begin{tabular}{ l c c c}
\hline
      ID & $\Theta$ & $N$ (m$^2\cdot s$) & $H_s$ (m) \\
\hline
  \verb"sw030" & $30^\circ$ & 0.720 & 4.63\\
  \verb"sw050" & $60^\circ$ & 0.719 & 4.6 \\
  \verb"sw170" & $180^\circ$ & 0.714 &  4.74\\
  \verb"sw230" & $240^\circ$ & 0.721 & 4.67\\
  \verb"sw330" & $330^\circ$ & 0.722 & 4.79\\
\hline
\end{tabular}
\label{table1}
\end{table}

\subsection{Self-similarity features of swell}
\begin{figure}
 \noindent\includegraphics[width=20pc]{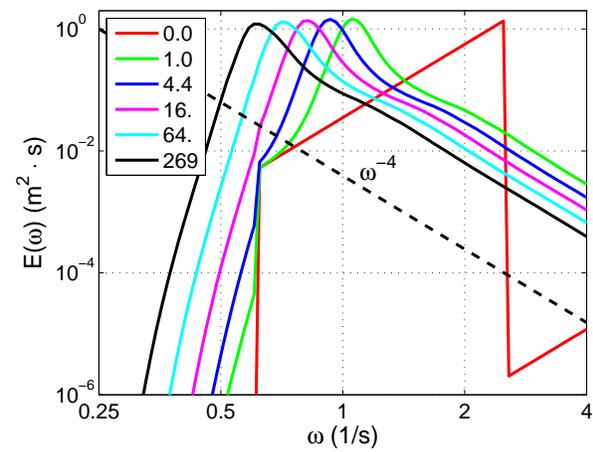}
 \caption{Frequency spectra at different times (legend, in hours) for the case sw330 ($\Theta=330^\circ$).}
 \label{bsifig1}
 \end{figure}

Evolution of swell spectra with time is shown in fig.~\ref{bsifig1} for the case \verb"sw330" of Table~\ref{table1}. The example shows a strong tendency to self-similar shaping of wave spectra. This remarkable feature has been demonstrated and discussed for swell in previous works \citep{BPRZ2005,Benoit_Gagnaire_2007,GagnaireEtal2010} for special parameters that provided relatively fast evolution of  rather short and unrealistically high waves. In our simulations we start with the mean wave period about $3$ seconds that corresponds to the end of calculations of \citet[][see fig.~8 therein]{BPRZ2005}  and moderately high steepness $\mu\approx 0.15$ as defined by (\ref{def:steepness}). The initial step-like spectrum evolves very quickly and keeps a characteristic shape for less than $1$ hour. For 264 hours the spectral peak period reaches $11$ seconds (the corresponding wavelength $\lambda\approx 186$ meters) and wave steepness becomes $\mu=0.023$. The final significant wave height $H_s \approx 2.8$ meters is essentially less than its initial value $4.8$ meters. The solution parameters can be considered as typical ones for ocean swell.

\begin{figure}
 \noindent\includegraphics[width=20pc]{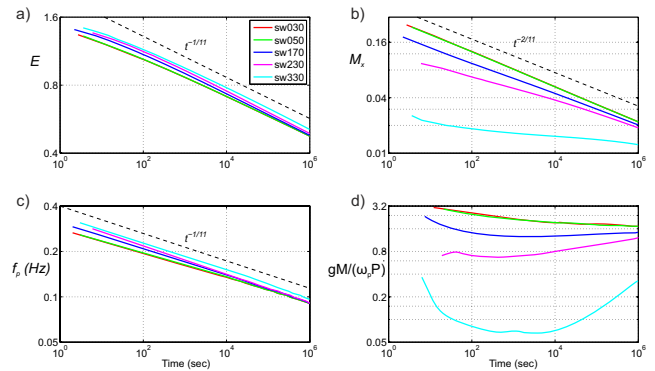}
 \caption{Evolution of wave parameters for runs  of Table~\ref{table1} (in legend):\emph{ a)} -- total energy $E$;\emph{ b)} -total wave momentum $M$; \emph{c)} -- frequency $f_p=\omega_p/(2\pi)$ of the energy spectra peak; \emph{d)} -- estimate of parameter of anisotropy in the Kolmogorov-Zakharov solution (\ref{eq:Kolmogorov_aniso}). Dashed lines show asymptotic power laws (\ref{eq:EnMom1},\ref{eq:downshift})}
 \label{bsifig2}
 \end{figure}

Dependence of key wave parameters on time is shown in fig.~\ref{bsifig2} for different runs of the series of Table~\ref{table1}. Power-law dependencies of self-similar solutions (\ref{eq:swelltau},\ref{eq:swellchi},\ref{eq:EnMom1}-\ref{eq:downshift}) are shown by dashed lines. In fig.~\ref{bsifig2}{\emph{a,b}} total wave energy $E$ and the spectral peak frequency $\omega_p$ show good correspondence to power laws of the self-similar solutions (\ref{eq:DurSS}). All the run curves are approaching the theoretical dependencies in a similar way. By contrast, power-law decay of the $x-$component of wave momentum $M_x$ depends essentially on angular spreading of initial wave spectra. While for narrow spreading (runs \verb"sw030" and \verb"sw050") there is no visible deviation from the law $t^{-1/11}$, wide-angle cases show these deviations definitely. The `almost isotropic' solution for \verb"sw330" is tending quite slowly to the theoretical dependency in terms of wave momentum $M_x$ (\ref{eq:EnMom2}). One can treat this transitional behavior by peculiarities of wave spectra relaxation from the `almost isotropic' state to an inherent distribution with a pronounced anisotropy.

A simple quantitative estimate of the `degree of anisotropy'  is given in fig.~\ref{bsifig2}{\emph{d}}. Evolution of dimensionless parameter of anisotropy in terms of the approximate Kolmogorov-Zakharov solution (\ref{eq:Kolmogorov_aniso}) by \citet{Katz_Kontor74}  is shown for all the cases of Table~\ref{table1}. We introduce parameter of anisotropy $A$ as follows
\begin{linenomath*}
\begin{equation}\label{def:A}
  A=\frac{gM}{\omega_p P}.
\end{equation}
\end{linenomath*}
where total energy flux $P$ (energy flux at $\omega\to \infty$) can be estimated  from evolution of total energy
\begin{linenomath*}
\begin{equation}\label{def:TotFluxE}
  P=-\frac{dE}{dt}.
\end{equation}
\end{linenomath*}
Similarly, total wave momentum
\begin{linenomath*}
\begin{equation}\label{def:WaveMomentum}
{\mathbf K}=\int \kv N(\kv)  d\kv
\end{equation}
\end{linenomath*}
provides an estimate of its flux in $x$-direction
\begin{linenomath*}
\begin{equation}\label{def:TotFluxM}
   M_x=-\frac{dK_x}{dt}.
\end{equation}
\end{linenomath*}
Spectral peak frequency $\omega_p$ has been used for the definition of `degree of anisotropy' $A$ (\ref{def:A}). Different scenarios are seen  in fig.~\ref{bsifig2}{\emph{d}} depending on angular spreading of wave spectra. Nevertheless, a general tendency to  a universal behavior at very large times (more than $10^6$ seconds) looks quite plausible.

\begin{figure}
 \noindent\includegraphics[width=20pc]{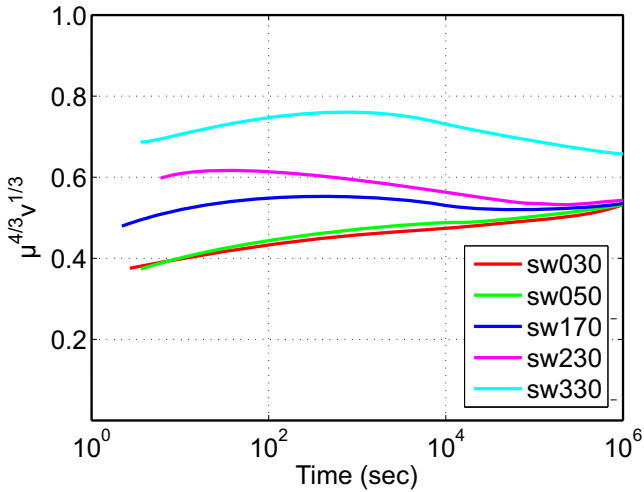}
 \caption{Evolution of the left-hand side of the invariant (\ref{eq:univ}) $(\mu^4\nu)^{1/3}$  for runs  of Table~\ref{table1} (in legend). }
 \label{bsifig3}
 \end{figure}

\begin{figure}
 \noindent\includegraphics[width=20pc]{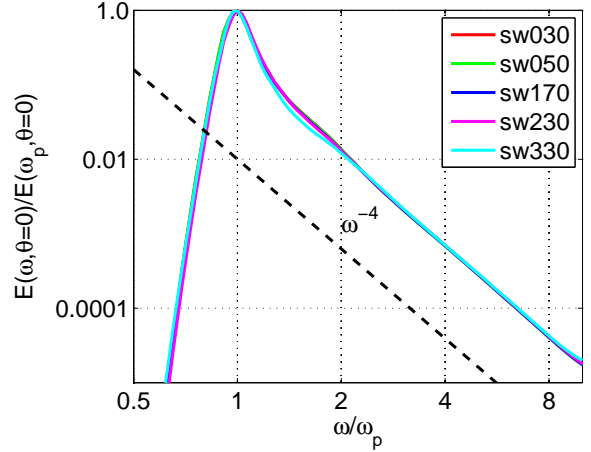}
 \caption{Normalized frequency  spectra for direction $\theta=0^{\circ}$ at $t=1000000$ seconds for runs  of Table~\ref{table1} (in legend). }
 \label{bsifig4}
 \end{figure}

Similar dispersion of different runs (different anisotropy of initial distributions) is seen in fig.~\ref{bsifig3} when tracing the invariant of the self-similar solutions (\ref{eq:univ}). One million seconds of the swell duration appear to be insufficient to present numerical arguments for the universality of the invariant (\ref{eq:univ}) for the self-similar solutions (\ref{eq:DurSS}). There is very likely a limit of the value at larger times. This limit is a bit less (by approximately $10\%$) than one for growing wind seas $\alpha_0\approx 0.7$. Again, the `almost isotropic' solution shows its stronger departure from the rest of the series. The differences are better seen in angular distributions rather than in normalized spectral shapes (fig.~\ref{bsifig4}) when we are trying to check  self-similarity features of the solutions in the spirit of \citet{BPRZ2005,Benoit_Gagnaire_2007}.

\subsection{Spectra angular spreading and its universality}
Angular spreading of the swell is an important issue of our consideration. Despite of dramatic difference of the runs  in integral characteristics of the swell anisotropy (e.g. figures \ref{bsifig2}{\em b,d}) the resulting spectral distributions still show clearly pronounced features of universality. This is of importance in the context  of remarks in sect.~2.2: does the only eigenfunction (or, more prudently, very few eigenmodes) survive in course of the swell evolution?

\begin{figure}
 \noindent\includegraphics[width=20pc]{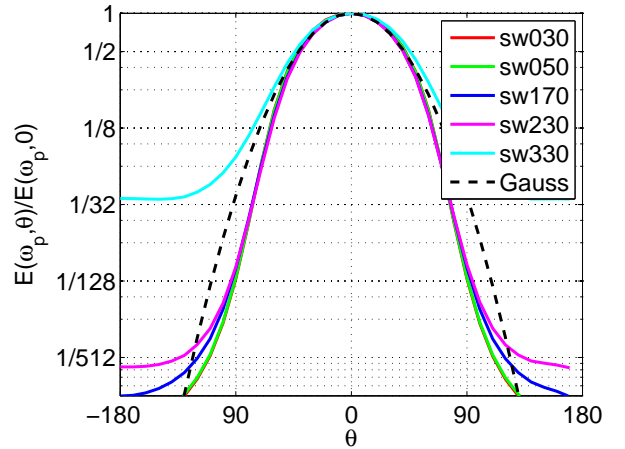}
 \caption{Normalized dependence of wave spectra on angle at peak frequency $\omega_p$ at $t=1000000$ seconds for runs  of Table~\ref{table1} (in legend). Gaussian distribution with dispersion $\sigma_\Theta=35^\circ$ is shown by dashed curve for reference.}
 \label{bsifig5}
 \end{figure}

 Normalized sections of spectra at the  peak frequency $\omega_p$ are shown in fig.~\ref{bsifig5} for the series runs at $t=10^6$ seconds. `The almost isotropic' run \verb"sw330" shows relatively high pedestal of about $3\%$ of maximal value while other series have a background less than $0.2\%$. At the same time, the core of all distributions is quite close to a gaussian shape \begin{linenomath*}
 \[
y= \exp\left(-\frac{\theta^2}{2 \sigma^2}\right)
\]
\end{linenomath*}
with half-width $\sigma = 35^\circ$.

\begin{figure}
 \noindent\includegraphics[width=20pc]{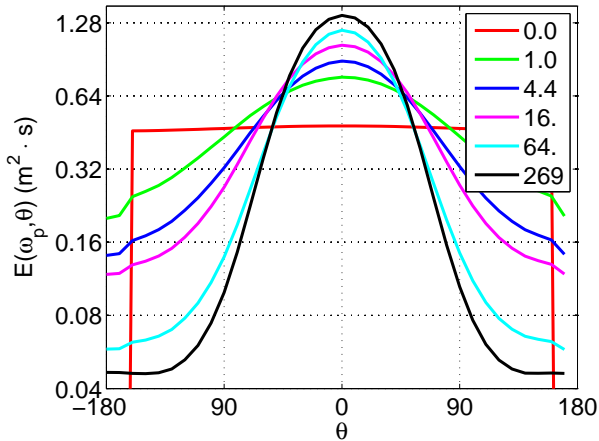}
 \caption{Angular distributions at the spectra peak frequency at different times (legend, in hours) for the case \texttt{sw330}.}
 \label{bsifig6}
 \end{figure}

Evolution of angular spreading in time is shown in fig.~\ref{bsifig6} for `the almost isotropic' run \verb"sw330" in absolute values. The sharpening of the angular distribution is accompanied by the peak increase of frequency-angle spectrum while the spectrum magnitude (averaged in angle) is slightly decaying with time (cf. fig.~\ref{bsifig1}).

\begin{figure}
 \noindent\includegraphics[width=20pc]{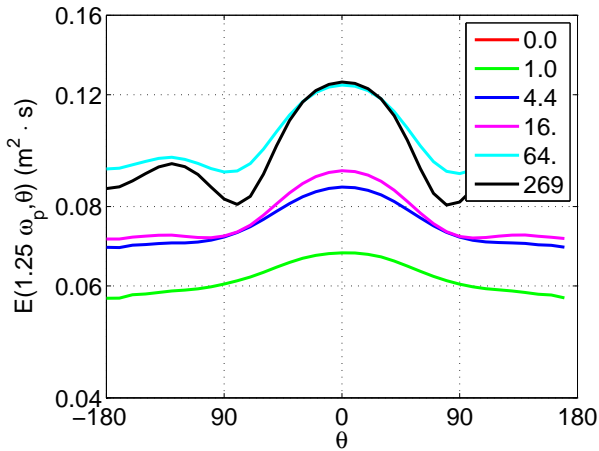}
 \caption{Angular distributions at frequency $1.25\omega_p$ at different times (legend, in hours) for the case \texttt{sw330}. Spectrum at $t=0$ is beyond of $y$-axis lower limit.}
 \label{bsifig7}
 \end{figure}

Angular spreading at higher frequency $1.25\omega$ in fig.~\ref{bsifig7}, first,  illustrates a rapid saturation of high-frequency range: energy level at this frequency increases by 5 orders in magnitude for $1$ hour only. Further evolution leads to pronounced sharpening of the distribution. The frequency $1.25\omega$ is a characteristic one in the Discrete Interaction Approximation (DIA) for the collision integral $S_{nl}$ which is used extensively in research and forecasting models of sea waves \citep{Hass_Hass85}. Thus, fig.~\ref{bsifig7} likely warns possible problems in simulations of wave-wave interactions with the DIA: the numerical scheme should have rather fine distribution for resolving peculiar features of spectral functions. At large time this figure shows formation of side lobes: maxima of spectral densities for oblique counter-propagating harmonics. Such behavior of wave spectra is reported in many papers \citep[e.g.][]{PRZ2003,BottemaVledder2008} and is usually discussed as an effect of wind input or  as a transitional effect. Our  simulations and the theoretical background of sect.~2.1 propose an alternative treatment of the effect in terms of properties of general stationary solutions (\ref{eq:KZgeneral}) for the kinetic equation (\ref{eq:Kinfull}). Fig.~\ref{bsifig7} shows clearly the presence of the second angular harmonics that can be predicted within the formal procedure of \citet{PRZ2003,PRZ2004}. Such treatment is not fully correct in the vicinity of the spectral peak but is still looks plausible. The second (and higher) harmonics of (\ref{eq:KZgeneral}) is strongly decaying with frequency  \citep[e.g.][]{Zakharov2010Scr} and their accurate account requires special analysis. In the next section we present  more evidences of their presence in the swell solutions.

 \subsection{Swell spectra vs KZ solutions}
\begin{figure}
 \noindent\includegraphics[width=10pc]{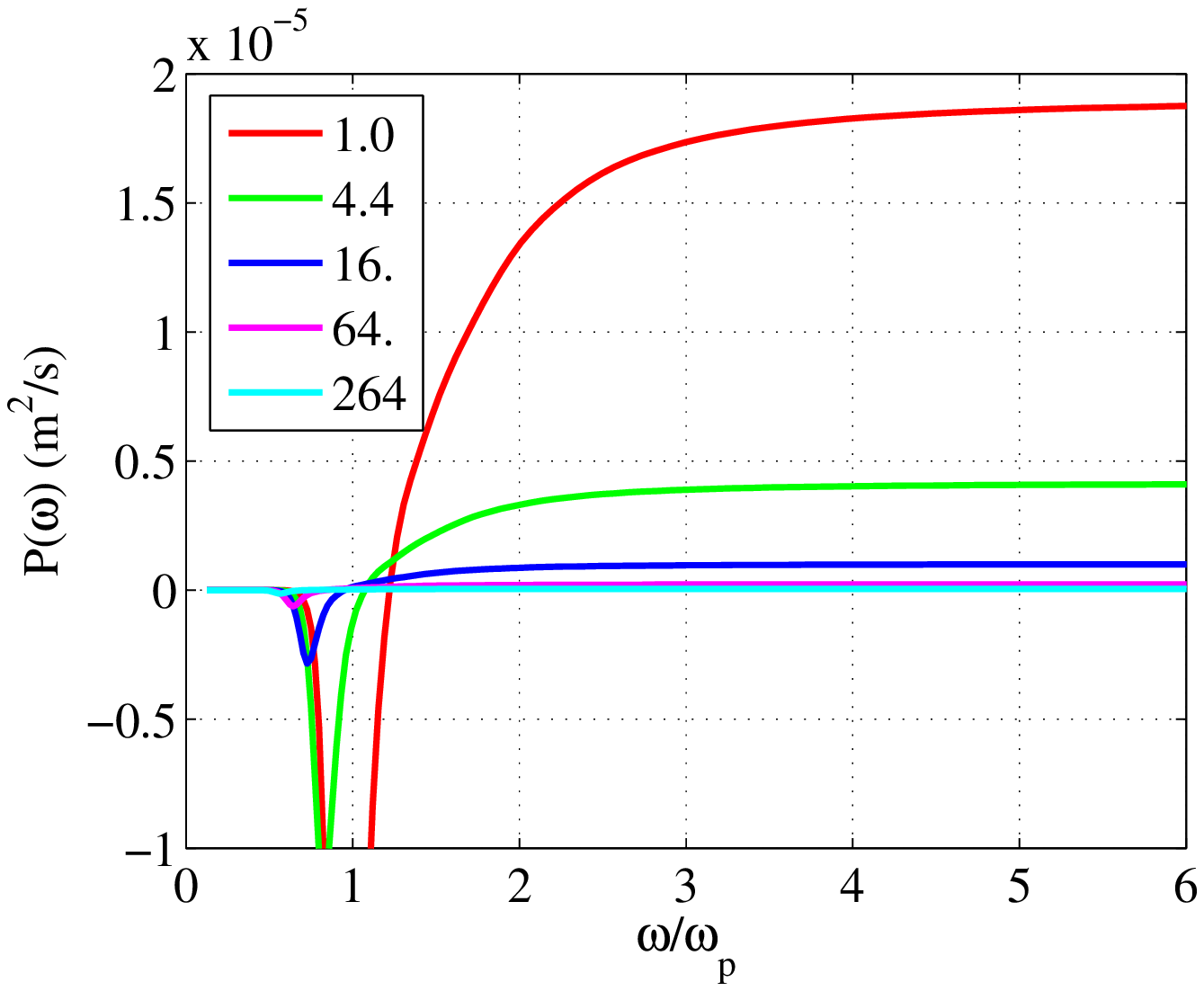}
\includegraphics[width=10pc]{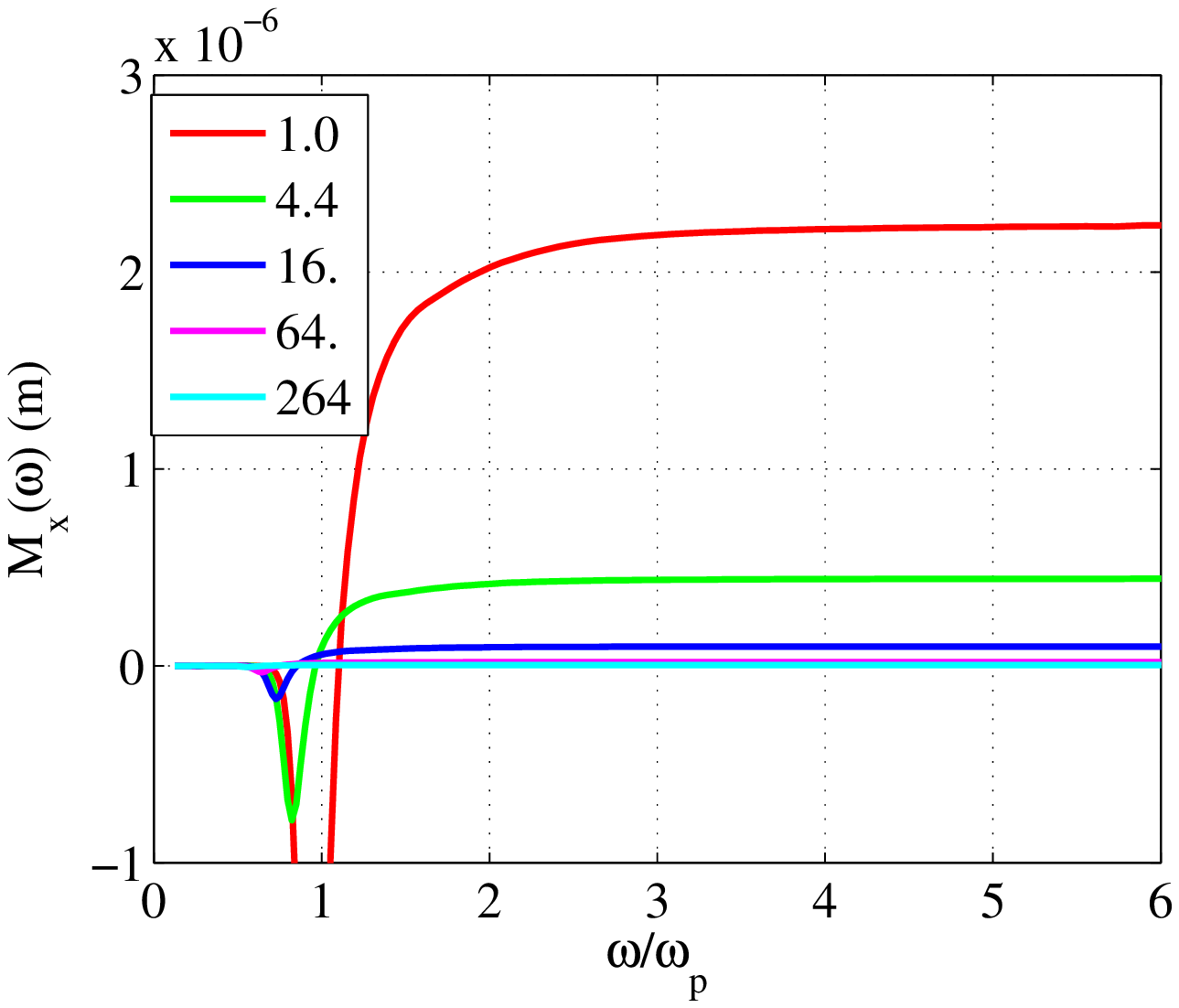}
 \caption{Left -- spectral fluxes of energy, right -- of the $x-$ component of wave momentum $M_x$ at different times (legend, in hours) for the case sw170.}
 \label{bsifig8}
 \end{figure}

 Very slow evolution of swell solutions in our simulations provides a chance to check relevance of the classic Kolmogorov-Zakharov solutions (\ref{eq:Kolmogorov_direct}-\ref{eq:KZgeneral}) to the problem under study. The key feature of the swell solution from the theoretical viewpoint is its `hybrid' \citep[in words of][]{BPRZ2005} nature: inverse cascading determines evolution of wave spectral peak and its downshifting while the direct cascading  occurs at frequencies slightly (approximately 20\%) above the peak. This hybrid nature is illustrated by fig.~\ref{bsifig8} for energy and wave momentum fluxes. In order to avoid ambiguity in treatment of the simulation results within the weak turbulence theory we will not detail this hybrid nature of swell solutions. Thus, general solution (\ref{eq:KZgeneral}) in the form
 \begin{linenomath*}
 \[
 E(\omega,\theta)== \frac{P^{1/3}g^{4/3}}{\omega^{4}}G(0,gM/(\omega P),\theta)
 \]
 \end{linenomath*}
 and its approximate explicit version (\ref{eq:Kolmogorov_aniso}) by \citet{Katz_Kontor71} will be used below for describing the direct cascading of energy and momentum at high (as compared to $\omega_p$) frequency.

Positive fluxes $P$ and $M$ decays with time in good agreement with power-law dependencies (\ref{eq:EnMom1}) and have rather low variations in  relatively wide frequency range $3\omega_p<\omega < 7\omega_p$ in fig.~\ref{bsifig8}. This domain of quasi-constant fluxes can be used for verification of results of the weak turbulence theory.

\begin{figure}
 \noindent\includegraphics[width=10pc]{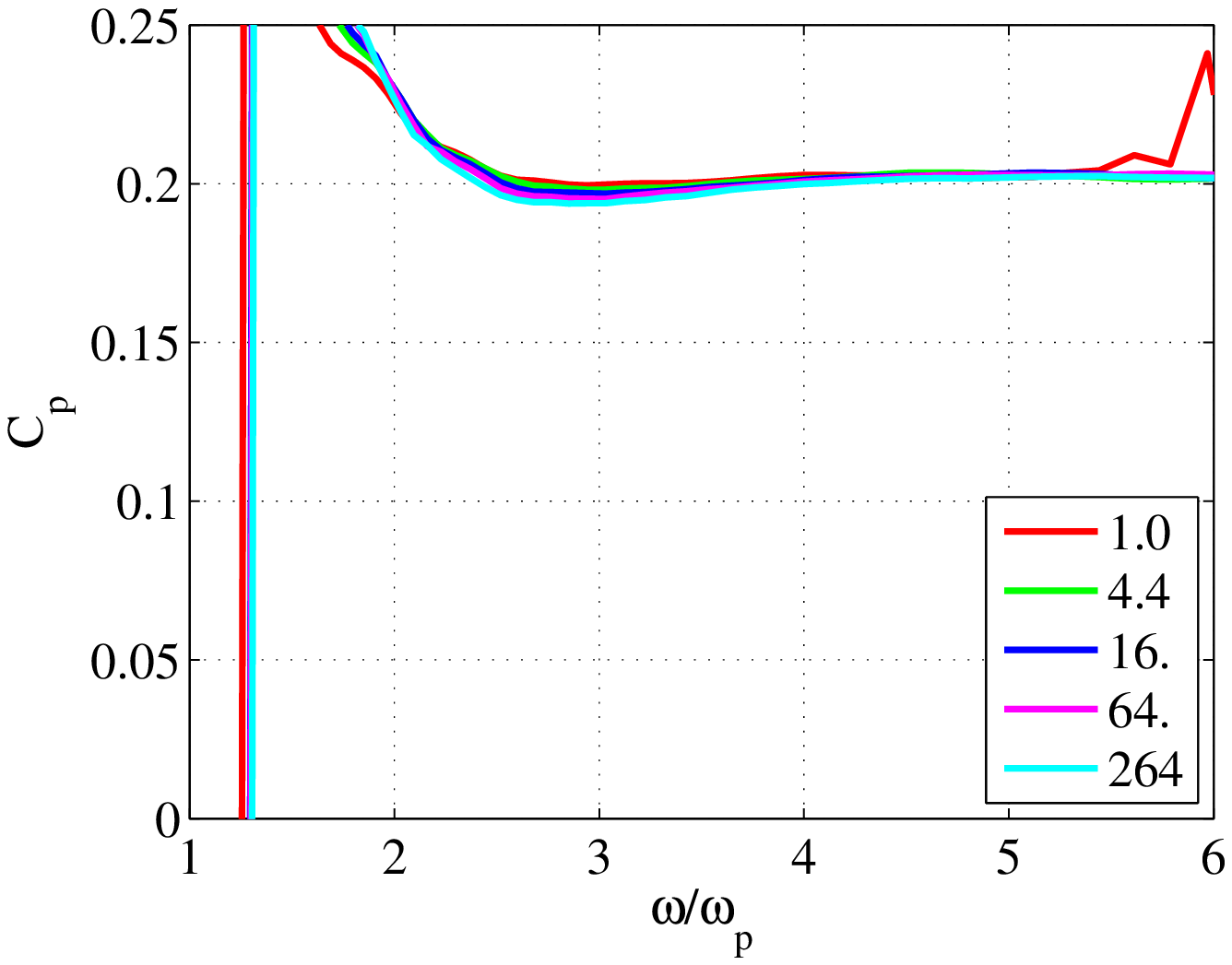}
\includegraphics[width=10pc]{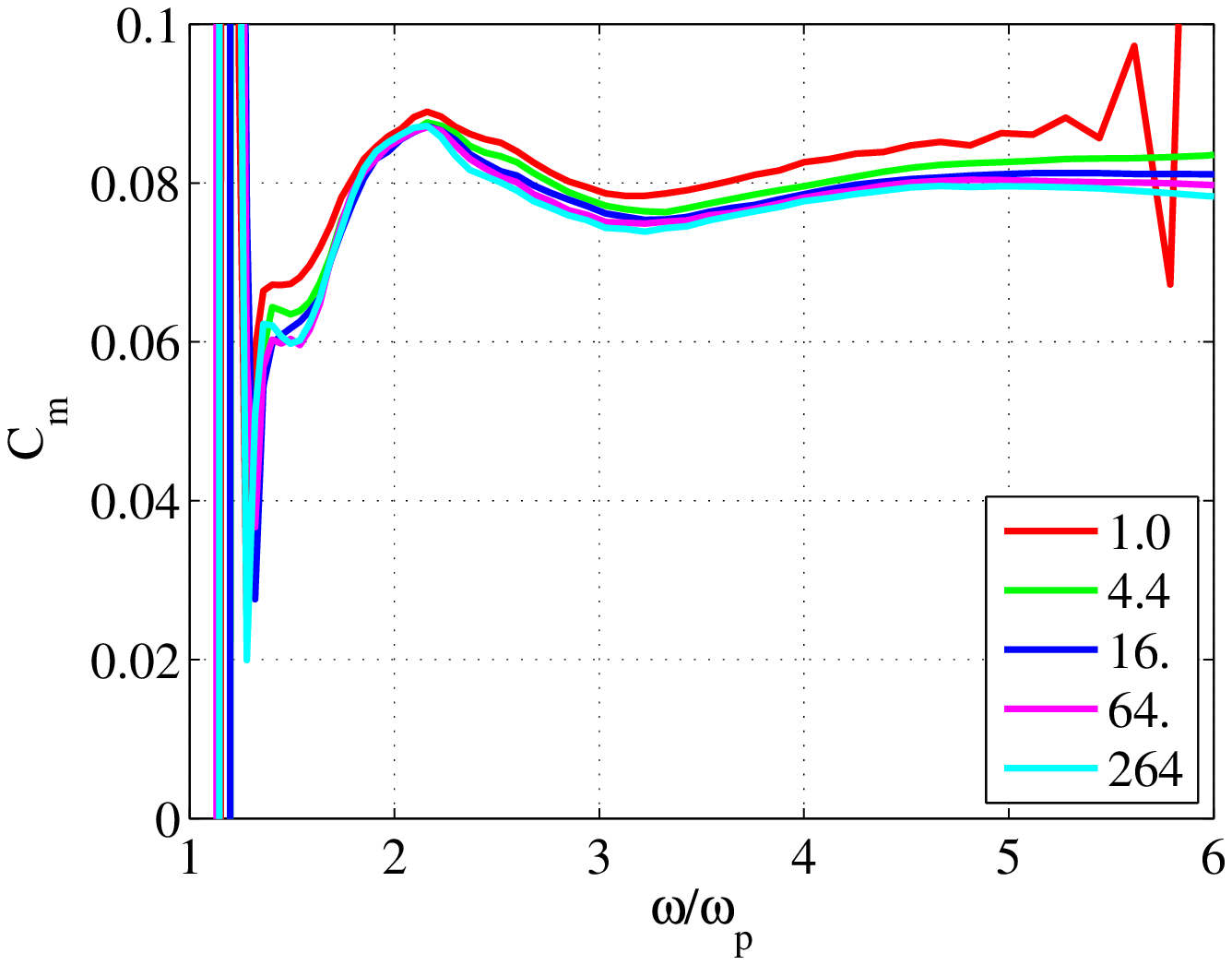}
 \caption{Left -- estimate of the first Kolmogorov constant $C_p$, right -- estimate of the second Kolmogorov constant $C_m$ for the approximate anisotropic KZ solution (\ref{eq:Kolmogorov_aniso}). Time in hours is given in legend for the case sw170.}
 \label{bsifig9}
 \end{figure}

The first and the second Kolmogorov's constants can be easily estimated using the approximate solution (\ref{eq:Kolmogorov_aniso}) from combinations of along- and counter-propagating spectral densities as follows
\begin{linenomath*}
\begin{eqnarray}\label{eq:2Cp2Cm}
  C_p=\frac{\omega^4\left(E(\omega,0)+E(\omega,\pi)\right)}{4g^{4/3}P^{1/3}}\\
  C_m=\frac{\omega^5P^{2/3}\left(E(\omega,0)-E(\omega,\pi)\right)}{4g^{7/3}M}.
\end{eqnarray}
\end{linenomath*}
The corresponding values  are shown in fig.~\ref{bsifig9}. The found estimates of Kolmogorov's constants $C_p\approx 0.21\pm 0.01$ and $C_m\approx 0.08\pm 0.02$ are consistent with previous results \citep{Lavrenov2002Banff,PRZ2003,BPRZ2005}. Note, that  the cited papers used different definitions of $C_p,\,C_m$. Here we follow one of \citet{Zakharov2010Scr} (see eqs.~\ref{eq:Kolmogorov_direct}-\ref{eq:Kolmogorov_aniso}).

Estimate of the second Kolmogorov constant $C_m$ requires special comments as far as its correctness can be affected by higher-order corrections (higher angular harmonics) of the solution (\ref{eq:Kolmogorov_aniso}). We see a possible effect of these corrections in relatively strong variations of estimates for the range $\omega/\omega_p < 4$ in fig.~\ref{bsifig9}{\em b}.

 The consistency of the estimates of the Kolmogorov's constants for the swell solutions does not mean a good fit of the approximate weakly anisotropic solution (\ref{eq:Kolmogorov_aniso}) to the results of our simulations. At the same time, the robustness of the estimates provides a good benchmark for the balance of spectral fluxes and energy level in the wave spectra.

 \section{Discussion. Simulations  for monitoring  ocean swell}
\subsection{Swell attenuation within the kinetic equation}
Results of the above consideration are two-fold. First, we illustrate the swell properties assuming the statistical approach for random water wave field (the Hasselmann equation) to be valid. We see that our simulations follow the basics of the weak turbulence theory quite well: deviations from self-similar asymptotics are generally small and  have physical explanation. The second aspect comes from this fact of consistency of theory and simulations: we see an additional support of our approach for simulating the long time swell evolution. Thus, the discussion of possible implications of our results to the problem of observation and monitoring ocean swell seems to be logical close of the work.

\begin{figure}
 \noindent\includegraphics[width=20pc]{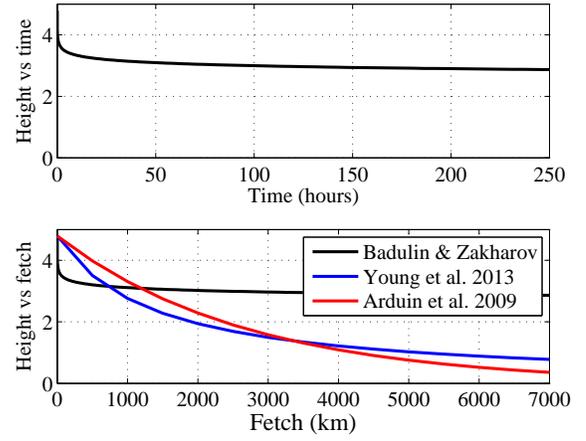}
 \caption{Top -- dependence of significant wave height $H_s$ on time  for the case \texttt{sw330}. Bottom -- attenuation of swell within different models \citet{Ardhuinetal2009,YoungBabZig2013jpo} and within one of this paper (see ledend). Results of duration-limited simulations are recasted into dependencies on fetch by simple transformation (\ref{eq:d2f}).}
 \label{bsifig10}
 \end{figure}

Dependence of wave height on time is shown in upper panel of fig.~\ref{bsifig10} for the `almost isotropic' run \texttt{sw330}. Strong down up to $30\%$ of initial value occurs at relatively short time of about one day. Essential part of the wave energy leakage corresponds to this transitional stage  at the very beginning of swell evolution when swell is tending very rapidly to a self-similar asymptotics. Afterwards the decay becomes much slower following the power-law dependence for the self-similar solutions (\ref{eq:EnMom1}) fairly well.

For comparison with other models and available observations the duration-limited simulations have been recasted into dependencies on fetch through the simplest time-to-fetch transformation \citep[e.g.][]{HwangWang2004,Hwang2006}:
\begin{linenomath*}
\begin{equation}\label{eq:d2f}
  x(s)=\int_0^s C_g(\omega_p(t)) dt.
\end{equation}
\end{linenomath*}
The equivalent fetch is estimated as a distance covered by wave guide travelling with the group velocity of the spectral peak component. As seen in fig.~\ref{bsifig10} our model predicts an abrupt drop of wave height at distances shorter than $1000$ km. In their turn, two quasi-linear models by \citet{Ardhuinetal2009} and \citet{Babanin2006turb} predict gradual decay of swell up to $10\%$ of initial wave height at distances $7000$ km where our model shows qualitatively opposite weak attenuation.

It should be noted that our model describes attenuation of the ocean swell `on its own' due to wave-wave interactions without any external effects. Thus, the  effect of an abrupt drop of wave amplitude at short time (fetch)  should be taken into consideration above all others when discussing possible application of our results to swell observations and physical treatment of the experimental results.

\subsection{Case study: swell under light wind}
An accurate account of weak (as compared to the models of swell decay mentioned above) swell attenuation due to the effect of wave-wave interactions deserves a fuller explanation. Deficiency of the today understanding of swell physics and, in particular, swell attenuation can be illustrated by the results of our special case study. The setup of the case followed literally the one described in sect.~3.1  with a minor difference: the effect of light wind has been added.  The wind input parameterization by \citet{Snyder81} is used as a conventional one for the today wind-wave modelling. The experiment mimics an effect of wind that changes its direction with inertial period about $16$ hours (inertial oscillations in the mid-latitude atmospheric boundary layer). The wind speed $5$ m/s was approximately $3$ times lower than phase speed of the spectral peak component at the start of the run (swell period $T_p\approx 10$ seconds corresponds to phase speed higher than $15$ m/s) and $5$ times lower at the end of the run.
 \begin{figure}
 \noindent\includegraphics[width=20pc]{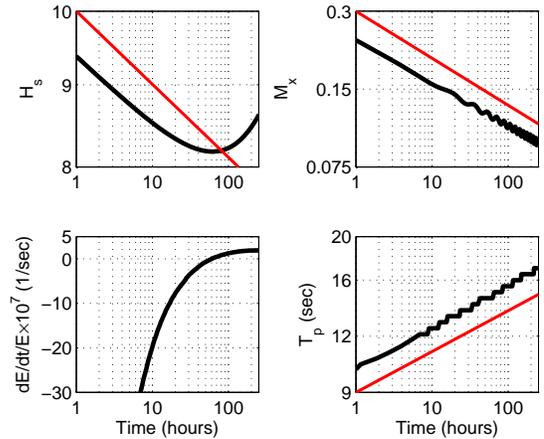}
 \caption{Results of simulations of swell under action of weak ($5$ m/s) rotating wind. Attenuation gives place to weak growth of swell after approximately 2 days of evolution. While dependence of wave height on time (a) is monotonous, wave momentum decay is accompanied by weak oscillations (b). Steps in peak wave period are from the simulation grid, no interpolation has been made.}
 \label{bsifig11}
 \end{figure}

As seen from fig.~\ref{bsifig11} the swell decaying at initial stage started to grow after approximately 2 days of evolution when its period was about $15$ seconds and wavelength exceeded $200$ meters. While wave height $H_s$ is growing (fig.~\ref{bsifig11}{\em a}) wave period $T_p$  continue to follow the self-similar law $t^{1/11}$ (\ref{eq:downshift}) fairly well. All the dependencies in fig.~\ref{bsifig11} are monotonous except one of the wave momentum $M_x$ ($T_p$ in fig.~\ref{bsifig11}{\em d} is not interpolated and is step-like because of numerical grid). The rotating wind is affecting the swell spectrum anisotropy but is feeding continuously wave energy. Wave-wave interactions are redistributing effectively energy of waves generated at high frequencies ($3$ or more times higher than peak frequency) and force swell to grow. This effect of the short wave absorption by swell has been discussed both for experimental data \citep{Young2006} and results of simulations within the Hasselmann equation and weakly nonlinear dynamical equations for water waves \citep{BKRZ2008Brest}. Manifestations of the effect in global wave datasets \citep[see][and http://www.sail.msk.ru/atlas/]{GulevGrig2003,GulevGrig2004} have been analyzed and presented in terms of `magic relations' (\ref{eq:magic}) as well \citep{BadulinGrigorieva2012}. Recent attempts to re-track swell from satellite data and, thus, to estimate attenuation rates also show essential portion of `negative dissipation rates' \citep[in words of][]{Jiang2016}. Most of the strange  dissipation rates, in authors opinion, are `not statistically significant'. At the same time, the distribution pattern of these rates correlates fairly well with results of analysis of the Voluntary Observing Ship (VOS) data \citep[cf. figs~4,7 of][and fig.~7 of \citet{Jiang2016}]{BadulinGrigorieva2012}.

Thus, the presented case study shows deficiency of the today models of swell attenuation where the role of inherently nonlinear evolution of swell spectra is completely ignored. Swell is an effective `devourer' of short wind waves \citep{BKRZ2008Brest} and it should be taken into account for the swell modelling.

\section{Conclusions}
We presented results  of sea swell simulations within the framework of the kinetic equation for water waves (the Hasselmann equation) and treated these properties within the paradigm of the theory of weak turbulence. A series of numerical experiments (duration-limited setup, WRT algorithm) has been carried out in order to outline features of wave spectra in a range of scales usually associated with ocean swell, i.e. wavelengths larger than $100$ meters and duration of propagation up to $10^6$ seconds ($\approx 11.5$ days). It should be stressed that the exact collision integral $S_{nl}$ (nonlinear transfer term) has been used in all the calculations. Alternative, mostly operational approaches, like DIA (Discrete Approximation Approach) can corrupt the results qualitatively.

Fix the key results of the study
\begin{enumerate}
\item First, the classic Kolmogorov-Zakharov (KZ) isotropic and weakly anisotropic solutions for direct and inverse cascades are shown to be relevant to slowly evolving sea swell solutions. Estimates of the corresponding KZ constants are found to agree well with previous works. Thus, KZ solutions can be used as a benchmark for advanced approaches in the swell studies;

  \item A strong tendency to self-similar asymptotics is observed. These asymptotics are shown to be insensitive to initial conditions. In particular, universal angular distributions of wave spectra at large times have been obtained for both narrow (appr. $30^\circ$) and almost isotropic initial spectra. The universality of the spectral shaping can be treated as an effect of mode selection when very few of eigenmodes of the boundary problem determines the system evolution. The inherent features of wave-wave interactions are responsible for this universality making effect of initial conditions insignificant. Initial conditions  can affect rates of the tendency to this asymptotic universal state. Generally, we observe self-similar swell development in a background which is far from self-similar state;

  \item   We show that an inherent peculiarity of the Hasselmann equation, energy leakage, can also be considered as a mechanism of the sea swell attenuation. This mechanism is not accounted for by the today models of sea swell. In the meantime, the energy decay rates of sea swell in the numerical experiments, generally, do not contradict to results of swell observations. Moreover, ability of wave-wave interactions to redistribute energy and wave momentum over the whole range of wave scales can explain  `negative attenuation rates' found in satellite data.
\end{enumerate}

The simulations of ocean swell within the duration-limited setup are just the very first step. Development of swell in space (fetch-limited setup) adds the effect of wave dispersion that can modify some results essentially.

----------------------------------------------------------------

\begin{acknowledgments}
Authors are thankful for the support of the Russian Science Foundation grant No. 14-05-00479.  Authors are indebted to Vladimir Geogjaev for discussions and comments.
\end{acknowledgments}


\end{article}

\end{document}